\documentclass[aps,prb,showpacs,twocolumn]{revtex4-1}
\usepackage{graphics}
\usepackage[dvips]{color}
\usepackage[T1]{fontenc}
\usepackage[latin1]{inputenc}
\usepackage{graphicx}
\usepackage{amssymb}
\usepackage{rotating}
\usepackage{epsfig}

\input epsf.sty

\newcommand{\ie}{{\em i.e.},\ }
\newcommand{\eg}{{\em e.g.},\ }
\newcommand{\etal}{{\em et al.}\ }

\newcommand{\mueof}{\mu_{{}_{\mbox{\tiny EOF}}}}
\newcommand{\gDPD}{\gamma_{{}_{\mbox{\tiny DPD}}}}
\newcommand{\kBT}{k_{{}_{\mbox{\tiny B}}}T}
\newcommand{\Pe}{\mbox{Pe}}
\newcommand{\Rey}{\mbox{Re}}

\newcommand{\dir}{{.}}

\begin{document}

\title{Separation of chiral particles in micro- or nanofluidic channels}
\author{Sebastian Meinhardt${}^1$, 
        Jens Smiatek${}^2$, 
        Ralf Eichhorn${}^3$, 
        and Friederike Schmid${}^1$}
\affiliation{
${}^1$ Institut f\"ur Physik, JGU Mainz, D-55099 Mainz, Germany \\
${}^2$ Institut f\"ur Physikalische Chemie, Westf\"alische Wilhelms-Universit\"at M\"unster, D-48149 M\"unster, Germany \\
${}^3$ Nordita, Royal Institute of Technology and Stockholm University, Roslagstullsbacken 23, SE-106 91 Stockholm, Sweden} 

\begin{abstract} 
We propose a method to separate enantiomers in microfluidic or nanofluidic
channels. It requires flow profiles which break chiral symmetry and have
regions with high local shear. Such profiles can be generated in channels
confined by walls with different hydrodynamic boundary conditions (\eg slip
lengths). Due to a nonlinear hydrodynamic effect, particles with different
chirality migrate at different speed and can be separated. The mechanism is
demonstrated by computer simulations. We investigate the influence of thermal
fluctuations (\ie the P\'eclet number) and show that the effect disappears in the
linear response regime.  The details of the microscopic flow are important and
determine which volume forces are necessary to achieve separation.
\end{abstract}


\maketitle

Enantiomers -- molecules that are not identical with their mirror image -- are
omnipresent in living organisms. Many biologically active molecules as well as
about half of the drugs in use today are enantiomers.  The pharmacological
activity of drugs strongly depends on their chirality, to the point that one
enantiomer may have a therapeutic effect, while its mirror image is toxic.
Therefore, the separation of enantiomers by chirality is of great importance
for basic science as well as for technical applications such as drug and
pesticide development or in food industry\cite{rekoske, guebitz}. 

In practice, chiral separation is mostly done by chemical methods, \eg by
introducing chiral selector molecules in conventional separation setups such as
liquid chromatography or capillary electrophoresis\cite{rekoske,guebitz,ward}
or in miniaturized microfluidic chips\cite{belder, nagl}. In recent years,
however, there has been increasing interest in developing physical separation
methods that do not require specific chiral recognition between 
molecules\cite{welch}. One of the first practical proposals due to de Gennes was to
separate chiral crystals by letting them slide down an inclined plane\cite{degennes}, 
where they would be slightly deflected from the axis of
maximal slope in directions that depend on the chirality. Speer \etal
generalized this concept of using potentials for separation and showed that
chiral particles can be forced to drift in opposite directions by a combination
of potentials that are periodic in time and space\cite{speer}. An alternative
idea was recently put forward by Spivak and Andreev who suggested to use
photoinduced drift in a gas buffer for chiral separation of small molecules\cite{spivak}. 

A particularly attractive approach is to use microfluidic flows for separation.
The term microfluidics refers to a set of rapidly evolving technologies for
preparing, controlling and analyzing minute amounts of liquids or gases, which
are believed to have an enormous technological potential in many areas like
pharmaceuticals, biotechnology, public health, or agriculture\cite{ouellette}.
A number of suggestions have recently been made how physical properties of
microfluidic flows could be exploited to separate enantiomers. Kostur \etal
proposed to introduce flow vortices which would trap particles of different
chirality at different places in space\cite{kostur}. One of us has designed a
strategy to separate enantiomers by taking advantage of different
translation-rotation couplings of stereoisomers in inhomogeneous flow\cite{eichhorn1, eichhorn2}.  

In these studies\cite{kostur, eichhorn1, eichhorn2}, hydrodynamic interactions
were neglected. Particles were assumed to undergo Brownian motion in an
externally imposed flow, without influencing the flow in return. However,
hydrodynamic interactions are known to play a decisive role for the motion of
microscopic particles in fluids\cite{dhont,brenner}. For example, chiral particles in
shear flow migrate in the vorticity direction due to hydrodynamics\cite{kim,
makino1}. Experimentally, this effect was verified for a wide range of Reynolds
numbers, ranging from macroscopic\cite{chen} to the millimeter\cite{makino2}
and micrometer scale\cite{marcos}. The direction of migration depends on the
Reynolds number, but the effect itself is universal, and it has been suggested
that it could be used to separate chiral particles\cite{marcos}.

In the present letter, we propose a method to achieve such a separation on the
micro- and nanoscale in microfluidic channels. The main idea is to
create an asymmetric flow profile with high-shear regions that drive particles
of different chirality to different regions in the channel, where they migrate
at different speed. This separates them while travelling through the channel.
To generate asymmetric flows, we propose to exploit a particular property of
micro- and nanoflows: The substrate-dependent slip at surfaces.  On the
macroscale, solid surfaces can usually be taken to impose
''no-slip'' boundary conditions on adjacent fluids, {\em i.e.}, the fluid
particles close to the surfaces move with the same velocity as the
surface.  On the micro- and nanometer scale, this is not necessarily
true, and particle slip may become significant, especially on hydrophobic
surfaces\cite{pit,neto}: The no-slip boundary condition has to be replaced by
a ''partial-slip'' boundary condition
$\delta_B \: \: \partial_{{\bf n}} v_{\parallel}|_{{\bf r}_B}
  =  v_{\parallel}|_{{\bf r}_B}$,
where $v_{\parallel}$ denotes the tangential component of the velocity and
$\partial_{{\bf n}} v_{\parallel}$ its spatial derivative normal to the
surface evaluated at the position ${\bf r}_B$ of the hydrodynamic boundary.
The slip length parameter $\delta_B$ depends on surface characteristics like
the roughness, the hydrophobicity etc. It can be tuned, e.g., by 
patterning the surfaces\cite{joseph}.  

Alternatively, surface flow can be induced by electric fields on 
charged surfaces due to an electrokinetic effect called ''electroosmotic flow''
(EOF)\cite{hunter}: Such surfaces are covered by a double layer of 
oppositely charged ions, which experience a force in an external electric field
and drag the surrounding fluid along. On homogeneous surfaces, this results in 
an effective boundary condition $v_{\parallel} = - \mueof E_{\parallel}$
for the flow outside the double layer, where $E_{\parallel}$ is the tangential 
component of the electric field and $\mueof$ the electroosmotic mobility.
Typical values on PDMS (polydimethylsiloxane) are\cite{streek} 
$\mueof \sim 3 \cdot 10^{-4} \text{cm}^2/\text{Vs}$. In the presence of surface slip, the 
magnitude of the effect can be strongly enhanced according to\cite{muller,bouzigues,
smiatek1} $\mueof = - \mueof^0 \: (1 + \kappa \: \delta_B)$, where
$\kappa = \partial_{{\bf n}} \psi/\psi \large|_{{\bf r}_B}$ is a screening
parameter describing the decay of the electrostatic potential $\psi$ at the
surface, and $\mueof^0 = -\epsilon_r \: \psi({\bf r}_B)/\eta_s$ the
well-known Smoluchowski result\cite{hunter} for the electroosmotic mobility 
at sticky walls. On patterned surfaces, the EOF behavior becomes even richer\cite{belyaev}. 
The possibility of playing with slip lengths and surface
charges by functionalizing and patterning surfaces in micro- and nanochannels
enables one to adjust boundary conditions and generate asymmetric profiles\cite{bogunovic,note1}, 
which could be used for chiral separation.

We first demonstrate the potential of this approach at the example of a channel
with simple quadratic cross-section, whose four walls are neutral (no
EOF) and have different slip lengths. We have carried out dissipative particle
dynamics (DPD) simulations\cite{hoogerbrugge, espanol} of short helical
particles which are carried through such a channel by pressure-driven flow. We
show that the channel indeed separates the particles effectively, and
demonstrate that this is a hydrodynamic effect, which vanishes almost entirely
if hydrodynamics are switched off. Finally, we discuss how the effect depends
on the shear rate in the channel and the exact form of the shear profile.

\begin{figure}[tb]
\includegraphics[width=0.36\textwidth]{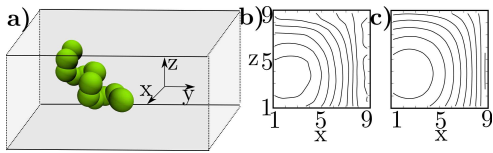}
\vspace*{-0.3cm}
\caption{\label{fig:model}
Illustration of the simulation model. A helical particle moves in a square
channel with asymmetric Poiseuille-like flow profile (a). Maps show the profiles
across the channel: (b) DPD flow profile; (c) Corresponding theoretical
profile calculated by numerical solution of the Stokes equation.  The
amplitudes depend on the applied pressure. 
}
\vspace*{-0.3cm}
\end{figure}

The basic geometric ingredients of the simulation model are sketched in Fig.\
\ref{fig:model}. In our DPD simulations, the fluid particles are transparent,
{\em i.e.}, they interact with each other only by dissipative interactions\cite{espanol}.  
In addition, they repel the wall and the monomers of the
chiral particles {\em via} repulsive Weeks-Chandler-Anderson (WCA) interactions\cite{WCA}. 
Partial slip boundary conditions are realized by introducing an
additional viscous coupling with the walls. This method allows one to realize
arbitrary slip lengths in a controlled manner\cite{smiatek2}.  Specifically,
the density of fluid particles in our simulations was chosen $\rho = 6
\sigma^{-3}$ and the DPD friction factor $\gDPD = 5 \tau \epsilon/\sigma^2$,
where the length unit $\sigma$ is the DPD and WCA interaction range,
the temperature unit is $\epsilon = \kBT$, and the time unit is $\tau = \sigma
\sqrt{m/\epsilon}$ with the particle mass $m$. The dynamic viscosity of this
fluid was measured with the periodic Poiseuille flow method\cite{backer},
giving $\eta_s = 1.35 \cdot \epsilon \tau/\sigma^3$. 

The fluid is confined to a square channel with cross section $10 \sigma \times
10 \sigma$, which corresponds to an accessible cross section $8 \sigma \times 8
\sigma$ and different slip lengths at all four walls\cite{note2}: $\delta_B^{(1)} = 11.23
\sigma$, $\delta_B^{(2)} = 5.54 \sigma$, $\delta_B^{(3)} = 1.58 \sigma$, and
$\delta_B^{(4)} = -0.17 \sigma$. For pressure driven flow, this
results in the profile shown in Fig.\ \ref{fig:model} (b), in good agreement
with the theoretical profile obtained by direct solution of the Stokes equation
(Fig.\ \ref{fig:model} c). The amplitude of the profile scales linearly with
the applied pressure. At volume force $f_y = 0.1 \epsilon/\sigma$, the flow
velocity in the channel varies between $0.2 \sigma/\tau$ and $3.6 \sigma/\tau$,
and the maximum shear rate in the high-shear region is $\dot{\gamma} \sim 1.0
\tau^{-1}$.  Mapping our model units to real SI units with a nanochannel
diameter $8 \sigma \sim 400 \text{nm}$ and the viscosity of water $\eta_s =
\eta_{\text{water}} \approx 1.\times 10^{-3} \text{Pa s}$ at room temperature
$T = \epsilon/k_{_B} \sim 300 \text{K}$, this places $\tau$ at about $20
\mu\text{s}$ and the maximum shear rate at $\dot{\gamma}_{\mbox{\tiny max}} \sim
0.5 \times 10^6 \: s^{-1}$, a value which is experimentally accessible in
nanochannels of this size\cite{kuang}. In a microchannel with diameter
$10 \mu\text{m}$, the same mapping gives $\tau \sim 3 s$ and 
$\dot{\gamma}_{\mbox{\tiny max}} \sim 0.3 s^{-1}$. 

The chiral particles are represented by short helices with two windings, made
of $N=13$ spherical monomers.  The helix geometry is characterized by three
parameters: The radius $R$, the helical pitch $p$, and the number of monomers
per turn $n$. Here we chose $R= 1 \sigma$, $p = 3 \sigma$, and $n=6$ (see Fig.\
\ref{fig:model}). This structure is stabilized by a set of stiff harmonic
potentials $V(X) = k_X/2 \: (X-X_0)^2$, where $X_0$ denotes the desired value
of the quantity $X$, and $X$ stands for (i) the bond lengths $d_i$
between neighbor monomers $i$ and $i+1$, (ii) the bending angles $\Phi_i$,
(iii) the torsional angles $\theta_i$, (iv) the distances $p_j$ between the
monomers $1,7$ and between $7,13$ (\ie the pitch), and (v) the angle $\Psi$
between the axes of the two windings. The spring parameters were set to
$k_d=k_p=500 \epsilon/\sigma^2$, $k_\phi=k_\theta=k_\Psi=500
\epsilon/\mbox{rad}^2$.  In our model fluid, this helix had a rotational
diffusion coefficient $D_R = 0.41 \cdot 10^{-2} \tau^{-1}$, a translational
diffusion coefficient $D_T = 0.018 \sigma^2/\tau$, and the hydrodynamic radius
$R_H = \kBT/6 \pi \eta_s D_T = 2.18 \sigma$.

Unless stated otherwise, the simulations were carried out at volume force $f_y
= 0.1 \epsilon/\sigma$ in a simulation box of length $50 \sigma$ and periodic
boundary conditions in the direction of flow $y$. The DPD time step was varied
between $\delta \tau = 0.001 \tau$ and $\delta \tau = 0.01 \tau$ to assess and
exclude finite time step artifacts.

\begin{figure}[tb]
\includegraphics[width=0.45\textwidth]{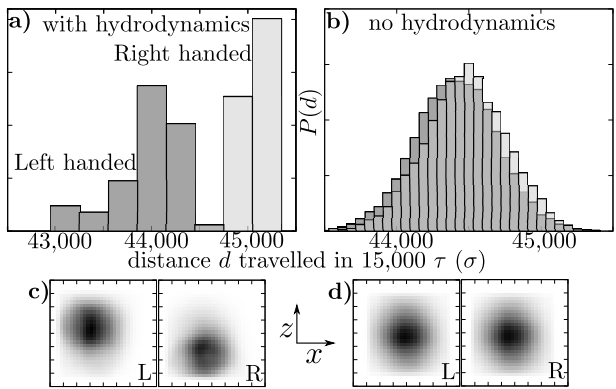}
\vspace*{-0.3cm}
\caption{\label{fig:separation}
Separation of chiral particles in the asymmetric channel at volume force $f_y =
0.1 \epsilon/\sigma$. a) Histogram of distances travelled in the time $t =
15.000 \tau$ for left handed helices (dark shaded) and right handed helices
(light shaded). b) Same as a) for the Langevin simulations without
hydrodynamics. c) Spatial distribution across the channel for left-handed helix
(left) and right-handed helix (right) for the simulations with hydrodynamics.
d) Same as c) for the Langevin simulations without hydrodynamics.
} 
\vspace*{-0.3cm}
\end{figure}

Our central simulation result is illustrated in Fig.\ \ref{fig:separation}. The
proposed separation strategy is clearly successful, helices of different
chirality travel through the channel with different speed and can be separated
along the channel (in the $y$ direction, Fig.\ \ref{fig:separation} a)). This
goes along with distinctly different spatial distributions across the channel
(in the $(x,z)$ plane, Fig.  \ref{fig:separation} c)). 

To assess the effect of hydrodynamic interactions, we have carried out Brownian
dynamics simulations of helices moving in the tabulated flow profile $\vec{V}(\vec{r})$
of Fig.~\ref{fig:model}~b). The equation of motion for monomers $i$
then reads $ m \ddot{\vec{r}}_i = \vec{f}_i - \zeta \: \big( \vec{v}_i -
\vec{V}(\vec{r}_i)\big) + \vec{\xi}_i$, where $\vec{f}_i$ is the force acting
on monomer $i$, $\vec{v}_i$ its velocity, and $\vec{\xi}_i$ an uncorrelated
Gaussian white noise which satisfies the fluctuation-dissipation theorem,
$\langle \xi_i^{\alpha}(t) \: \xi_j^{\beta}(t') \rangle = 2 \kBT \zeta \:
\delta_{ij} \: \delta_{\alpha \beta} \: \delta(t-t')$.  The friction
coefficient $\zeta$ was chosen such that the translational diffusion constant
in the Brownian dynamics simulations matches that of the DPD simulations, \ie
$\zeta = \kBT/N D_T = 4.27 m/\tau$.  The time step in these simulations was
chosen $\delta \tau = 0.0001 \tau$.  The results are shown in Fig.\
\ref{fig:separation} b) and d). In contrast to the results from the DPD
simulations, the distributions for the two enantiomers are almost identical
(Fig.  \ref{fig:separation} d)) and so are their velocities (Fig.\
\ref{fig:separation} b)). Thus, the separation in the DPD simulations is driven
by hydrodynamics.

\begin{figure}[tb]
\includegraphics[width=0.35\textwidth]{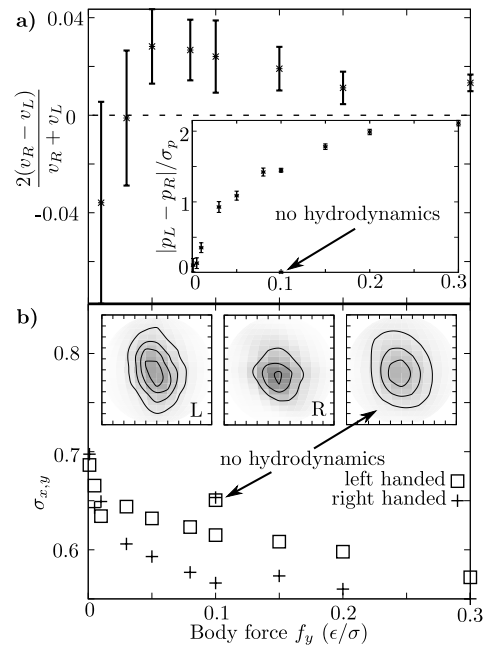}
\vspace*{-0.3cm}
\caption{\label{fig:variation}
Separation of chiral particles in the asymmetric channel as a function of
volume force $f_y$.  a) relative velocity difference for lefthanded and
righthanded particles. Inset: distance between means of the distributions $p_L$
and $p_R$ for lefthanded and righthanded particles in the cross-section of the
channel (the $(x,z)$-plane) divided by the mean standard deviation
$\sigma_p=(\sigma_{p_L}+\sigma_{p_R})/2$ of the two distributions.  b) Width
(standard deviation) $\sigma_{x,z}$ of the distribution $P(u_x,u_z)$ of helix
orientations $\vec{u}$ projected on the $(x,z)$-plane. The limiting value for
equidistribution on a sphere is $\sigma_{x,y}\sim 0.82$. Insets: Corresponding
distributions at $f_y = 0.1 \epsilon/\sigma$ for left-handed helices (L),
right-handed helices (R), and helices without hydrodynamic interactions.
}
\vspace*{-0.5cm}
\end{figure}

The underlying mechanism, the shear-induced drift of chiral particles in the
vorticity direction, is a nonlinear effect in bulk solution, which is forbidden
in the linear response regime\cite{makino1}. The shear flow must first orient
the particles before it can induce drift.  One would expect the separation to
fail as one reaches the linear response regime. The crucial quantity in this
picture is the dimensionless ratio of the shear rate $\dot{\gamma}$ and the
rotational diffusion constant, the P\'eclet number $\Pe = \dot{\gamma}/D_R$.
To study this dependence on \Pe\ in more detail, we have varied the
average shear rate systematically by varying the volume force $f_y$ on the
particles.  Fig.\ \ref{fig:variation} a) shows the relative velocity difference
between helices of different chirality as a function of $f_y$.  As expected, it
vanishes within the error for small forces. The cross-sectional distributions
of the left-handed and right-handed particles approach each other continuously
as one reduces the force (Fig.\ 3 a, inset), and separation is no longer
possible.  These results can be correlated with the orientation distribution of
the helices.  Fig.\ 3 b) shows that helices are mostly oriented in the flow
direction.  A certain amount of orientation is also observed in the simulations
without hydrodynamics, probably due to the mechanisms described in Ref.\
\onlinecite{eichhorn1}.  It accounts for the very small shift between the
distance distributions for left-handed and right-handed particles in Fig. 2 b).
However, the dominating effect is clearly due to hydrodynamics. With decreasing
$f_y$, the orientation distribution broadens until one recovers the value
observed without hydrodynamics at low forces, $f_y\lesssim
0.02\epsilon/\sigma$.  Taking into account that $f_y \sim 0.1\epsilon/\sigma$
corresponds to average P\'eclet numbers of order 120, Fig.\ \ref{fig:variation}
suggests that P\'eclet numbers of around 30 are neccessary to achieve
separation in our channels.

\begin{figure}[tb]
\includegraphics[width=0.35\textwidth]{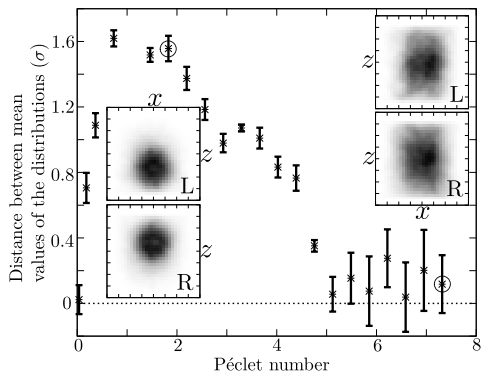}
\vspace*{-0.3cm}
\caption{\label{fig:shear}
Spatial separation of chiral particles in a symmetric square channel
with constant shear in $x$-direction as a function of P\'eclet number. 
The insets show explicit distributions at the encircled values.
}
\vspace*{-0.3cm}
\end{figure}

The specific threshold value depends on the details of the flow profile.  For
comparison, we have also studied square channels with constant shear in the
$x$-direction by enforcing antiparallel fluid velocities at the opposing
$(y,z)$-boundaries.  Technically, such channels can be realized by inducing EOF
of different amplitudes on two opposing sides of the channel combined with
full-slip boundary conditions on the remaining two sides. Since they
have symmetric flow profiles, they will not propel chiral particles with distinct
(average) migration velocities and thus cannot be used to separate particles in
the $y$ direction along the channel. Particles of different chirality 
migrate at equal speed in contrast to the asymmetric profiles discussed
above, but they may still occupy different regions in the
$(x,z)$-plane, and this effect also depends on the P\'eclet number as shown in
Fig.\ \ref{fig:shear}. As one increases the P\'eclet number, the distributions
of left-handed and right-handed particles in the $(x,z)$-plane 
move apart. This sets in at much earlier P\'eclet numbers than in the
asymmetric channel ($\Pe \sim 0.5$). However, a second effect comes into play
at higher P\'eclet numbers, which causes the distributions to move together
again and merge again at $\Pe \sim 5$.  The reasons for this unexpected
behavior are not yet clear. Most likely, it is due to the onset of inertial
effects at higher Reynolds numbers \Rey, which push the Helices towards the
center of the channel\cite{carlo}. This effect should thus disappear
in real nano- or microchannels, where \Rey\ is small, whereas it is of order
$\Rey \sim 10$ in our DPD simulations.

To summarize, we have demonstrated that it is possible to separate enantiomers
on microscopic scales in asymmetric flows without explicit chiral agents.  By
comparing simulations with and without hydrodynamic interactions, we have shown
that the separation is dominated by hydrodynamic mechanisms.  They can be
exploited to force particles of different chirality into different regions of
the channel where they migrate at different speed. We have
demonstrated the effect for simple, not optimized, square geometries, and we
have shown that it is highly susceptible to the flow profile. Other
(e.g., rectangular) channel geometries can presumably be designed where
the separation will be even more efficient.

This work was funded in part by the VW foundation and the German Science
Foundation within SFB-TR6. We thank Peter Reimann for useful discussions.


\begin{thebibliography}{66}
\bibitem{note1} When exploiting EOF, one must ensure that
the chiral particles are not attracted by the charged walls and that the
channel diameter is large compared to the thickness of the charged double layer.
($\sim 1$ nm in physiological buffers).
\bibitem{note2} Even though the effective slip length is negative, the fluid
velocities in the surface region are of course always positive.
\end{thebibliography}
\end{document}